# MULTI-DIMENSIONAL CUSTOMIZATION MODELLING BASED ON METAGRAPH FOR SAAS MULTI-TENANT APPLICATIONS

Ashraf A. Shahin[1, 2]

[1]College of Computer and Information Sciences,
Al Imam Mohammad Ibn Saud Islamic University (IMSIU)
Riyadh, Kingdom of Saudi Arabia
ashraf_shahen@ccis.imamu.edu.sa
[2]Department of Computer and Information Sciences,
Institute of Statistical Studies & Research, Cairo University, Egypt

## ABSTRACT

*Software as a Service (SaaS) is a new software delivery model in which pre-built applications are delivered to customers as a service. SaaS providers aim to attract a large number of tenants (users) with minimal system modifications to meet economics of scale. To achieve this aim, SaaS applications have to be customizable to meet requirements of each tenant. However, due to the rapid growing of the SaaS, SaaS applications could have thousands of tenants with a huge number of ways to customize applications. Modularizing such customizations still is a highly complex task. Additionally, due to the big variation of requirements for tenants, no single customization model is appropriate for all tenants. In this paper, we propose a multi-dimensional customization model based on metagraph. The proposed mode addresses the modelling variability among tenants, describes customizations and their relationships, and guarantees the correctness of SaaS customizations made by tenants.*

## 1. INTRODUCTION

Software as a Service (SaaS) is a cloud computing service model in which pre-built applications are delivered to customers as a service [1]. The main goal of the SaaS providers is attracting a significant number of tenants for their SaaS applications. However, the functionality and quality that individual customers require from a software application can differ [2]. This forces the SaaS providers to deploy multiple software applications customized for each set of users, which results in increasing cost of infrastructure and making it difficult to maintain and update. To overcome with this problem, the SaaS providers deploy software applications allow tenant-specific configuration and customization. For customization, some elements of an application need to be customized, including graphic user interfaces (GUI), workflow, service, and data models [3]. Several researches have attempted to support configuration and customization of these elements [1, 4, 5, 6, 7].

One of the most prominent approaches for achieving the tenant-specific configuration and customization is providing an application template with unspecified parts that can be customized by each tenant [7, 8]. These unspecified parts are called customization points of an application [3]. For each customization point, a set of components are provided to achieve variations in tenants' requirements. A tenant customizes each customization point by selecting components from the corresponding set or by defining new components.

However, modularizing such customizations still is a highly complex task. The complexity, in general, comes from the following sources. First, the SaaS applications could have hundreds of customization points with thousands of components for achieving variations in tenants' requirements. Second, the relationships and dependences of components are more complex. It is possible for some components to require other components or for some components to conflict with others (figure 1 shows the relationships of components from different customization points). Third, due to the big variation of requirements for the tenants, no single customizations model is appropriate for all tenants. To address this variation, the developers need to provide software applications with different modularizations.

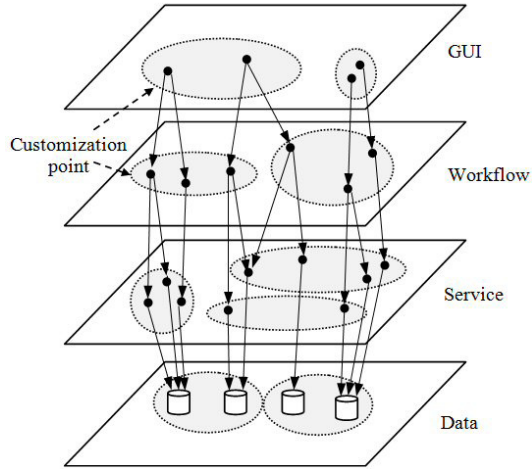

Figure 1. Relationships of components from different customization points

*Multi-dimensional separation of concerns (MDSOC)* is separation of concerns that allows developers to partition overlapping or interacting concerns in new and existing software products along multiple dimensions of decomposition and composition. *MDSOC* allows developers to identify new concerns, and new dimensions of concern, incrementally and simultaneously, at any time [9]. Another important property for *MDSOC* is *on-demand remodularization*, which is the ability to change the way in which a system is modularized without having to refactoring, reengineering, or other invasive change [10, 11]. A detailed description of MDSOC approach is available in [9].

In this paper, we apply MDSOC concepts to address the modelling variability among tenants. Applying MDSOC allows developers to provide independent dimension for each type of concerns and allows tenants to customize concerns that crosscut several customization points independently. For example, tenants that are concerned with security can customize the security concerns in security dimension only. This results in distinct customization models, which are easier to handle than one large model. On the other hand, we use metagraph to describe dependences between components of each concern. Metagraph is a graphical structure in which edges represent directed relationships between sets of elements [12]. Using metagraph tools, we can combine the separated models to validate the customization points' values made by tenants.

This paper is organized as the following. Section 2 discusses the related work in customization modelling. Section 3 describes our multi-dimensional customization model. Section 4 evaluates the proposed model. Section 5 concludes this paper.

## 2. RELATED WORK

Several researchers have tried to modularize customizations in SaaS applications [13, 14, 15]. In [13], the authors propose a method for modelling customization process based on Temporal

Logic of Actions, and propose a verification algorithm to check tenant's customization and to guarantee that each step in customization will not cause unpredictable influence on system and follow the related rules defined by SaaS provider. In [3], the authors propose a customization model based on metagraph to describe relationships between customization points and propose an algorithm to calculate related sets when one customization point is changed. In [14], the authors propose a multi-granularity customization relationship model that uses directed graph to describe relations in the customization process. The authors introduce a verification algorithm to guarantee the correctness of SaaS customization products that are obtained by the proposed model. In [15], the authors model the customization process using orthogonal variability model and propose a guided semi-automated customization based on mining existing tenants' customization.

In [16], the authors propose an approach for model-level customization management, which raises the customization management to the level of business services. They specify variation points using so-called specialization patterns, which have been originally developed to support task-driven specialization of application frameworks. In [17], the authors propose multi-level customization model for SaaS applications. To reduce duplication of customization, the proposed model supports customization sharing among different virtualized applications in a tenant area. The proposed model reduces repeated customization operations by allowing lower-level applications to inherit high-level applications' customizations. However, most of current research work has not addressed the modelling variability among tenants.

## 3. MULTI-DIMENSIONAL CUSTOMIZATION MODELLING BASED ON METAGRAPH

In this section, we begin by giving an overview of the basic features of metagraphs from [12]. This is followed by a description of the proposed model. Finally, we will propose two algorithms to validate customizations made by tenants.

### 3.1. Metagraphs

Metagraphs are graphical structures in which edges represent directed relationships between sets of elements. They extend both directed graphs (by allowing multiple elements in vertices) and hypergraphs (by including directionality in edges).

Definition 1 (metagraph): Given a finite generating set $X = \{x_i, i = 1 \ldots I\}$, a metagraph is an ordered pair $S = \langle X, E \rangle$, in which $E$ is set of edges $E = \{e_k, k = 1\ldots K\}$. Each edge is an ordered pair $e_k = \langle V_k, W_k \rangle$, in which $V_k \subseteq X$ is the invertex of the edge $e_k$ and $W_k \subseteq X$ is the outvertex. The coinput of any $x \in V_k$ is $V_k \backslash \{X\}$ and the cooutput of any $x \in W_k$ is $W_k \backslash \{X\}$. Also $V_k \cup W_k \neq \Phi$ for all $k$.

For example, the metagraph in Figure 2 can be represented as follows:

$S = \langle X, E \rangle$, where

$X = \{x_1, x_2, x_3, x_4, x_5, x_6\}$ is the generating set, and

$E = \{e_1, e_2, e_3\}$ is the set of edges where each edge can be specified as pair ( for example $e_1 = \langle \{x_1, x_2\}, \{x_3, x_4\} \rangle$ )

Invertex $(\langle \{x_1, x_2\}, \{x_3, x_4\} \rangle) = \{x_1, x_2\}$,

Outvertex $(\langle \{x_1, x_2\}, \{x_3, x_4\} \rangle) = \{x_3, x_4\}$,

Coinput $(x_1, \langle \{x_1, x_2\}, \{x_3, x_4\} \rangle) = \{x_2\}$,

Cooutput $(x_3, \langle \{x_1, x_2\}, \{x_3, x_4\} \rangle) = \{x_4\}$,

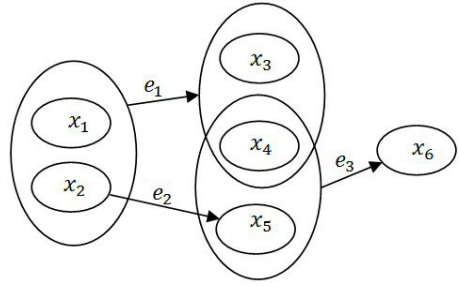

Figure 2. Metagraph example

Definition 2 (simple path): Given a metagraph S =⟨X , E⟩, a simple path from a source $x \in X$ to a target $x' \in X$ is a sequence of edges $h(x, x') = \langle e'_l, l = 1, \ldots, L \rangle$ such that $x \in V'_1$, $x' \in W'_L$, and $W'_i \cap V'_{i+1} \neq \Phi, \forall i = 1, \ldots, L-1$. The coinput of x is $(\cup_{l=1}^{L} V'_l \setminus \cup_{l=1}^{L} W'_l) \setminus \{x\}$ and the cooutput of x' is $\cup_{l=1}^{L} W'_l \setminus \{x'\}$.

The length of a simple path is the number of edges in the path; thus, the length of $h(x, x') = \langle e'_l, l = 1, \ldots, L \rangle$ is L. For example, in the metagraph of Figure 2, $\langle e_1, e_3 \rangle$ is a simple path from $x_2$ to $x_6$ with length 2. The coinput of $x_2$ is $\{x_1, x_5\}$, and the cooutput of $x_6$ is $\{x_3, x_4\}$.

Definition 3 (metapath): Given a metagraph $S = \langle X, E \rangle$, a metapath M(B,C) from a source $B \subseteq X$ to a target $C \subseteq X$ is a set of edges $E' \subseteq E$ such that (1) $e' \in E'$ is on a simple path from some element in B to some element in C, (2) $[\cup_{e'} V_{e'} \setminus \cup_{e'} W_{e'}] \subseteq B$, and (3) $C \subseteq \cup_{e'} W_{e'}$. For example, in Figure 2, one metapath from $x_1$ to $x_6$ is $M(\{x_1, x_2\}, \{x_6\}) = \{e_1, e_2, e_3\}$.

Definition 4 (adjacency matrix): the adjacency matrix A of a metagraph is a square matrix with one row and one column for each element in the generating set X. The ijth element of A, denoted $a_{ij}$, is a set of triples, one for each edge e connecting $x_i$ to $x_j$. Each triple is of the form $\langle CI_e, CO_e, e \rangle$ in which $CI_e$ is the coinput of $x_i$ in e and $CO_e$ is the cooutput of $x_j$ in e. For example, the adjacency matrix for the metagraph in Figure 2 is shown in Figure 3.

|       | $x_1$ | $x_2$ | $x_3$ | $x_4$ | $x_5$ | $x_6$ |
|-------|-------|-------|-------|-------|-------|-------|
| $x_1$ | $\phi$ | $\phi$ | $\{\langle\{x_2\},\{x_4\},e_1\rangle\}$ | $\{\langle\{x_2\},\{x_3\},e_1\rangle\}$ | $\phi$ | $\phi$ |
| $x_2$ | $\phi$ | $\phi$ | $\{\langle\{x_1\},\{x_4\},e_1\rangle\}$ | $\{\langle\{x_1\},\{x_3\},e_1\rangle\}$ | $\{\langle\phi,\phi,e_2\rangle\}$ | $\phi$ |
| $x_3$ | $\phi$ | $\phi$ | $\phi$ | $\phi$ | $\phi$ | $\phi$ |
| $x_4$ | $\phi$ | $\phi$ | $\phi$ | $\phi$ | $\phi$ | $\{\langle\{x_5\},\phi,e_3\rangle\}$ |
| $x_5$ | $\phi$ | $\phi$ | $\phi$ | $\phi$ | $\phi$ | $\{\langle\{x_4\},\phi,e_3\rangle\}$ |
| $x_6$ | $\phi$ | $\phi$ | $\phi$ | $\phi$ | $\phi$ | $\phi$ |

Figure 3. The adjacency matrix of the metagraph in figure 2

Definition 5 (closure matrix): the closure of adjacency matrix A, denoted $A^*$, represents all simple paths of any length in the metagraph. The ijth element of $A^*$, denoted $a^*_{ij}$, is a set of triples, one for each simple path $h(x_i, x_j)$ of any length connecting $x_i$ to $x_j$. The closure matrix of Figure 3 appears in Figure 4.

Definition 6: Given a generating set X and two metagraphs $S_1 = \langle X, E_1 \rangle$ and $S_2 = \langle X, E_2 \rangle$ with adjacency matrices $A_1$ and $A_2$ respectively, then the sum of the two adjacency matrices is the adjacency matrix of the metagraph $S_3 = \langle X, E_1 \cup E_2 \rangle$ with components $(A_1 + A_2)_{ij} = a^1_{ij} \cup a^2_{ij}$.

|     | $x_1$ | $x_2$ | $x_3$ | $x_4$ | $x_5$ | $x_6$ |
|-----|-------|-------|-------|-------|-------|-------|
| $x_1$ | $\phi$ | $\phi$ | $\{\langle\{x_2\},\{x_4\},e_1\rangle\}$ | $\{\langle\{x_2\},\{x_3\},e_1\rangle\}$ | $\phi$ | $\{\langle\{x_2,x_5\},\{x_3,x_4\},\langle e_1,e_3\rangle\rangle\}$ |
| $x_2$ | $\phi$ | $\phi$ | $\{\langle\{x_1\},\{x_4\},e_1\rangle\}$ | $\{\langle\{x_1\},\{x_3\},e_1\rangle\}$ | $\{\langle\phi,\phi,e_2\rangle\}$ | $\{\langle\{x_1,x_5\},\{x_3,x_4\},\langle e_1,e_3\rangle\rangle,\langle\{x_4\},\{x_5\},\langle e_2,e_3\rangle\rangle\}$ |
| $x_3$ | $\phi$ | $\phi$ | $\phi$ | $\phi$ | $\phi$ | $\phi$ |
| $x_4$ | $\phi$ | $\phi$ | $\phi$ | $\phi$ | $\phi$ | $\{\langle\{x_5\},\phi,e_3\rangle\}$ |
| $x_5$ | $\phi$ | $\phi$ | $\phi$ | $\phi$ | $\phi$ | $\{\langle\{x_4\},\phi,e_3\rangle\}$ |
| $x_6$ | $\phi$ | $\phi$ | $\phi$ | $\phi$ | $\phi$ | $\phi$ |

Figure 4. The closure of the adjacency matrix in Figure 3

Definition 7: A metagraph $S_1 = \langle X_1, E_1 \rangle$ is said to be a *sub-metagraph* (SMG) of another metagraph $S_2 = \langle X_2, E_2 \rangle$ if $X_1 \subseteq X_2$ and $E_1 \subseteq E_2$. A metagraph $S_1$ is an *input independent SMG* of a metagraph $S_2$ if every element of $S_1$ that is not a pure input is determined only by edges within $S_1$. A metagraph $S_1$ is an *output independent SMG* of a metagraph $S_2$ if every element of $S_1$ that is not a pure output is used only by edges within $S_1$. A metagraph $S_1$ is an *independent SMG* of a metagraph $S_2$ if it is both input independent and output independent.

## 3.2. The proposed model

Consider $CP = \{cp_i, i = 1, \ldots, I\}$ is the set of all customization points in SaaS application. For each customization point $cp_n \in CP$, there is a set of components $C_n = \{x_k, k = 1, \ldots, K\}$ provided by the SaaS application developer. Tenant customizes customization point $cp_n$ by selecting a subset of components $TC_n \subseteq C_n$.

Definition 8 (concern): Each concern that is important for tenants is represented by a metagraph $CN_i = \langle X_{cn_i}, E_{cn_i} \rangle$, where $X_{cn_i}$ is a set of all components that address this concern from all customization points' component sets (as shown in figure 5), and $E_{cn_i}$ is the relationships of these components.

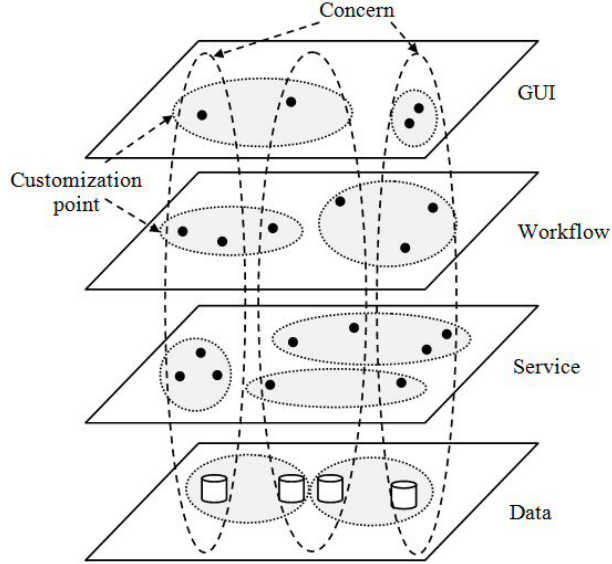

Figure 5. Concerns crosscut several customization points

In this paper, we concerned with "*required by*" relationship. Sometimes component requires more than one component. To express this state we add "*and*" vertex to the metagraph. For example, figure 6 shows that components $x_1$ and $x_2$ are required by component $x_4$ and components $x_2$ and $x_3$ are required by component $x_5$.

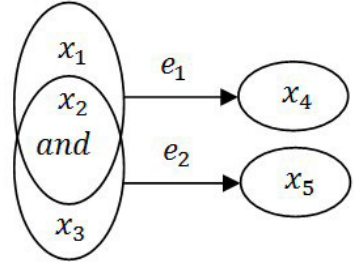

Figure 6. Metagraph example

Concerns are considered to overlap if their component sets are not disjoint. If there is a component address more than one concern, the required components can be different depending on the role played by this component in each concern.

Tenant customizes concern by selecting some components from its component set. The customized concern can be represented as a sub-metagraph $TCN_i \subseteq CN_i$. To guide tenants during customization process, closure matrix of each concern is generated to return all possible customization paths of this concern. Tenants review these paths and make their decisions.

Definition 9 (dimension of concerns): Dimension of concerns is a type of concerns that is important for tenants (e.g., performance, security) such that each dimension of concerns $D_d = \{cn_i, i = 1,..,I\}$ contains a set of concerns whose component sets are disjoint and each concern is an independent sub-metagraph. Dimension can be represented as a Metagraph $D_d = \langle X, E_d \rangle$, where $X$ is the set of all components in SaaS application, and $E_d = E_{cn_1} \cup E_{cn_2} \cup ... \cup E_{cn_n}$, $E_{cn_m}$ is a set of edges for one of the concerns belong to this dimension. The adjacency matrix of each dimension is the sum of all adjacency matrixes of its concerns.

Each dimension, has one concern called the *None* concern. All components in SaaS application that do not address any concern in this dimension address the *None* concern. This means that each component in the SaaS application will be located at exactly one concern in each dimension.

Definition 10 (SaaS application): A SaaS application is a set of dimensions $D = \{d_j, j = 1,..,J\}$, such that every concern is in exactly one dimension in $D$. A SaaS application can be represented as a metagraph $S = \langle X, E \rangle$, $X$ is the set of all components in SaaS application, and $E = E_1 \cup E_2 \cup ... \cup E_n$, $E_m$ is the set of edges for one of the dimensions in this application.

Definition 11 (customization operations): is what a tenant does to customize the SaaS application to meet his requirements. In this paper, we define two types of customization operations: add component, and delete component. Tenant customizes SaaS application by a sequence of these operations such that each operation moves SaaS application from a valid state to another valid state.

Definition 12 (customization model): customization for each tenant are represented as a set $TD = \{td_j, j = 1,..,J\}$, where $td_j$ is the set of customized concerns in dimension $d_j$. each tenant's customization is a sub-metagraph from SaaS application metagraph.

Tenants customize SaaS application by selecting dimensions that are important for them (e.g., performance, security), and select concerns from each dimension (e.g., authentication, authorization, and encryption in security dimension). Finally, tenants can customize each concern by specifying components from its component set.

## 3.3. Customization validation

Tenant customizes the SaaS application using a sequence of customization operations. To validate this sequence, we validate each operation based on the customizations that are constructed by the previous operations. If tenant requests to add new component to a specific concern in his customization, algorithm 1 shows how to validate and perform this operation. Algorithm 1 adds new components only if their requirements already exist in the tenant's customization. Algorithm 2 deletes components from the tenant's customization only if these components are not required by any component in the tenant's customization. Algorithm 2 deletes all edges that contain the deleted component in its outvertex without any cooutput. To minimize run-time and memory usage for the proposed algorithms, the only required rows or columns of adjacency matrixes are constructed.

<u>Algorithm 1 Add component</u>
1: FUNCTION add
2: IN: $CN = \langle X_{cn}, E_{cn} \rangle$, the concern metagraph from the SaaS application
3: IN: $TD = \langle X_{td}, E_{td} \rangle$, the tenant's customization metagraph.
4: IN: $CN_{td} = \langle X_{cn_{td}}, E_{cn_{td}} \rangle$, the customization of concern $CN$ from $TD$
5: IN: $x_i$, the component added to customization.
6: OUT: $v$, *valid* or *invalid*
7: BEGIN
8: Construct adjacency matrix column $x_i$ of $CN$ as $A_{cn_i}$.
9: for each $a_{x_j,x_i}$ in $A_{cn_i}$
10:   if $a_{x_j,x_i} \neq \phi$
11:     for each triple $\langle CI, CO, e \rangle$ in $a_{x_j,x_i}$
12:       if "and" $\in CI$
13:         if $(\{x_j\} \cup CI) \subseteq X_{td}$
14:           $X_{cn_{td}} = X_{cn_{td}} \cup \{x_i\} \cup (\{x_j\} \cup CI)$
15:           $E_{cn_{td}} = E_{cn_{td}} \cup \{e\}$
16:           return *valid*
17:         else if $(\{x_j\} \cup CI) \cap X_{td} \neq \phi$
18:           $X_{cn_{td}} = X_{cn_{td}} \cup \{x_i\} \cup \left((\{x_j\} \cup CI) \cap X_{td}\right)$
19:           $E_{cn_{td}} = E_{cn_{td}} \cup \{e\}$
20:           return *valid*
21: for each $a_{x_j,x_i}$ in $A_{cn_i}$
22:   if $a_{x_j,x_i} \neq \phi$
23:     return *invalid*
24: $X_{cn_{td}} = X_{cn_{td}} \cup \{x_i\}$
25: return *valid*

<u>Algorithm 2 Delete component</u>
1: FUNCTION delete
2: IN: $TD = \langle X_{td}, E_{td} \rangle$, the tenant's customization metagraph.
3: IN: $x_d$, the component deleted from customization.
4: OUT: $v$, *valid* or *invalid*
5: BEGIN
6: Construct adjacency matrix row $x_d$ of $TD$ as $R_{td_d}$.
7: for each $a_{x_d,x_i}$ in $R_{td_d}$
8:   if $a_{x_d,x_i} \neq \phi$
9:     return *invalid*
10: Construct adjacency matrix column $x_d$ of $TD$ as $C_{td_d}$.
11: for each $a_{x_j,x_d}$ in column $x_d$ in $C_{td_d}$
12:   if $a_{x_j,x_d} \neq \phi$

13:    for each triple $\langle CI, CO, e \rangle$ in $a_{x_j,x_d}$
14:        if $CO = \phi$
15:            for each concern $CN_{td} = \langle X_{cn_{td}}, E_{cn_{td}} \rangle$ in $TD$
16:                $E_{cn_{td}} = E_{cn_{td}} - \{e\}$
17:                $E_{td} = E_{td} - \{e\}$
18: for each concern $CN_{td} = \langle X_{cn_{td}}, E_{cn_{td}} \rangle$ in $TD$
19:    $X_{cn_{td}} = X_{cn_{td}} - \{x_d\}$
20: return *valid*

## 4. EVALUATION

There are a number of benefits for the proposed model. Some of these benefits have been explored below:

- Customization complexity: by applying MDSOC, the proposed model allows tenant to customize one concern at a time instead of customization point. Concerns cut across multiple customization point to collect components that are related to a specific area of interest. In other words, tenant customizes customization points incrementally. Which enables tenant easily customizes the customization points and easily understands their dependencies. On the other side, provider follows the same steps in developing. Instead of developing all components of a specific customization point, provider develops components of a specific concern.

- SaaS upgradeability: in some SaaS application, tenants may be encountering some issues when SaaS providers upgrade their applications. The proposed model allows providers to add new concerns and new dimensions at any time without having to reengineering existing ones. Providers can upgrade each concern independently because each concern is modeled as independent sub-metagraph and is unaffected by changes that may occur within its associated dimension. Using metagraph tools, providers can expect the effects of their upgrades on the tenants' customizations within each concern.

- Duplication of customization: the SaaS providers need to reduce duplication of customization to exploit the economies of scale [2]. Most of duplications of customization come from allowing tenants to define new components. This makes the SaaS providers face a broad spectrum of tenants' customizations and makes it difficult to drive commonalities amongst the customizations across tenants. By allowing providers to model a large number of components and enabling tenants to handle these components easily, providers can provide a wide range of components to reduce duplication of customization.

- Customization correctness: the proposed algorithms and metagraph tools had been used to ensure the correctness of customizations made by tenants.

However, with the proposed model, providers can face some difficulties in specifying disjoint concerns and their components in each dimension. We believe that, graph decomposition tools can help providers in specifying these concerns.

To evaluate the run-time performance of the proposed model and algorithms for differing application sizes, we had generated customization models with different sizes and random dependences. Figures 7 shows the average response time for each model size.

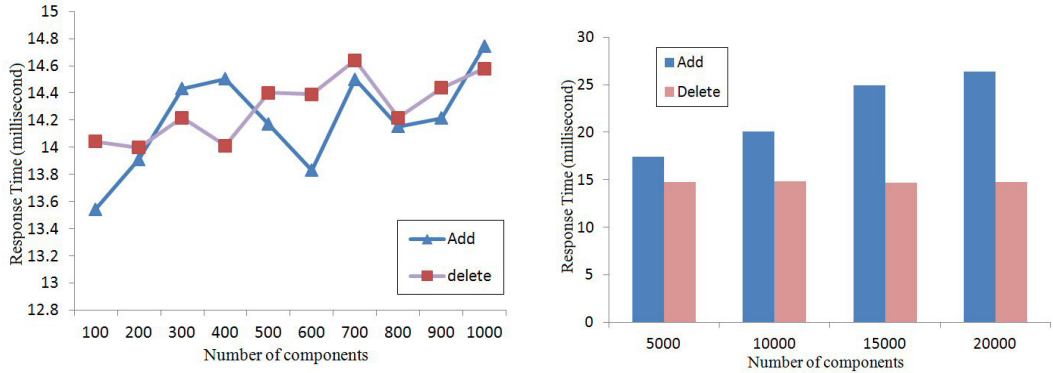
Figure 7. Response time for different application sizes

To evaluate the run-time performance of the proposed model and algorithms for varying numbers of users, we had generated model with size 500 components and had performed add and delete operations on the generated model with a large number of concurrent requests from users. The results are described in Figure 8.

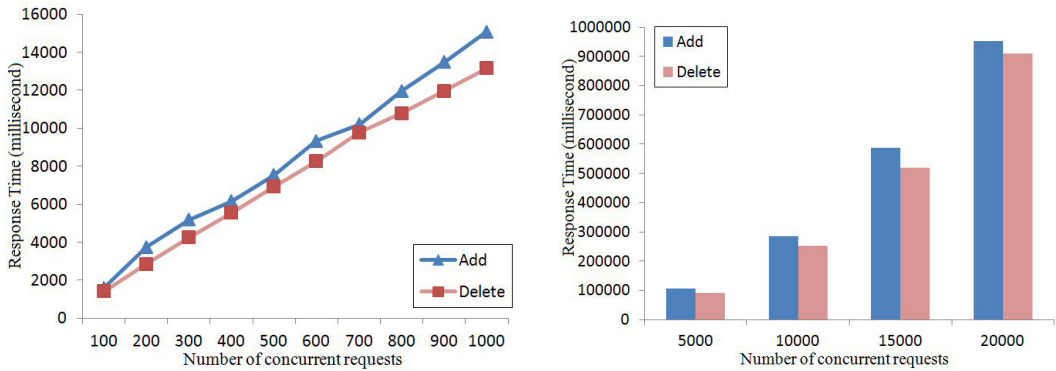
Figure 8. Response time for different numbers of concurrent requests

## 5. CONCLUSION

In this paper, we presented a multi-dimensional customization model. The proposed model applied the MDSOC concepts to address the modelling variability among tenants and described components' dependences using metagraph.

The main benefits of our model are: (i) simplifying customization process by allowing tenants to incrementally customize customization points; (ii) increasing SaaS upgradeability by allowing providers to add new concerns and new dimensions at any time without having to reengineering existing ones; and (iii) reducing duplication of customization by minimizing the needs for define new components. Additionally, we proposed algorithms that can judge whether tenant's customization action has satisfied all the dependency, and prevent the invalid action.

However, specifying disjoint concerns and their components in each dimension (which is beyond the scope of this paper) still is a challenge for the proposed model. We can use graph-decomposition tools to specify these concerns.

Finally, we can conclude that, applying MDSOC concepts to customization models addresses the modelling variability among tenants. Our future work will be to improve the proposed model to guide tenants during the customization process based on mining existing tenants' customizations.